\documentclass[a4paper]{article}
\usepackage[utf8]{inputenc} 
\usepackage{textcomp} 
\usepackage{graphicx}  
\usepackage{flafter}  

\usepackage{amsmath,amssymb,amsfonts,amsthm}  
\usepackage{bm}  
\usepackage{bbm}
\usepackage{mathtools}
\usepackage{enumitem}
\usepackage{authblk}
\mathtoolsset{showonlyrefs}
\usepackage[numbers]{natbib}

\usepackage{verbatim}
\usepackage{xcolor}

\usepackage[bookmarks,colorlinks,breaklinks]{hyperref}  
\definecolor{dullmagenta}{rgb}{0.4,0,0.4}   
\definecolor{darkblue}{rgb}{0,0,0.4}
\hypersetup{linkcolor=red,citecolor=blue,filecolor=dullmagenta,urlcolor=darkblue} 
\usepackage{memhfixc}  
\usepackage{pdfsync}  

\newtheorem{theorem}{Theorem}

\newtheorem{innercustomgeneric}{\customgenericname}
\providecommand{\customgenericname}{}
\newcommand{\newcustomtheorem}[2]{%
  \newenvironment{#1}[1]
  {%
   \renewcommand\customgenericname{#2}%
   \renewcommand\theinnercustomgeneric{##1}%
   \innercustomgeneric
  }
  {\endinnercustomgeneric}
}

\newcustomtheorem{customthm}{Theorem}
\newcustomtheorem{customlemma}{Lemma}

\usepackage{braket}

\def\tr{{ \rm tr}}

\def\cE{\mathcal E}
\def\cH{\mathcal H}

\renewcommand{\rho}{\varrho}

\def\id{\mathbbm{1}}

\newcommand{\proj}[1]{|#1\rangle\langle#1|}
\newcommand*{\cB}{\mathcal{B}}
\newcommand*{\cF}{\mathcal{F}}

\newcommand*{\cT}{\mathcal{T}}

\usepackage{epstopdf}

\usepackage{xspace}
\def\etal{\emph{et al.\xspace}}

\begin{document}

\title{Coherent state coding approaches the capacity of non-Gaussian bosonic channels}
\author{Stefan Huber and Robert K\"onig}
\affil{Institute for Advanced Study \& Zentrum Mathematik,\\ Technical University of Munich, 85748 Garching, Germany}
\date{\today}
\maketitle
\begin{abstract}
  The additivity problem asks if the use of entanglement can boost the information-carrying capacity of a given channel beyond what is achievable by coding with simple product states only. This has recently been shown not to be the case for phase-insensitive one-mode Gaussian channels, but remains unresolved in general. Here we consider two general classes of bosonic noise channels, which include phase-insensitive Gaussian channels as special cases: these are attenuators with general, potentially non-Gaussian environment states and classical noise channels with general probabilistic noise. We show that additivity violations, if existent, are rather minor for all these channels: the maximal gain in classical capacity is bounded by a constant independent of the input energy. Our proof shows that coding by simple classical modulation of coherent states is close to optimal.
\end{abstract}
\section{Introduction}

Communication -- be it over time (as storage in a memory) or over space (as transmission from a sender to a receiver) -- is one of the central primitives 
studied in information theory. A  channel
represents a general model for communication. 
With respect to communication, its arguably most  fundamental characteristic 
 is its classical capacity: the maximal number of 
bits or -- more precisely -- the maximal rate at which bits may be transmitted through the channel asymptotically in the limit of many  uses. This quantity is ubiquitous in information theory since it has both practical meaning and is interesting from a purely mathematical point of view. 

The classical capacity of quantum channels has been studied for decades~\cite{schumacherwestmore,holevocapacity,Caves_1994}, but is not understood in general. Not only is it very hard to find the optimal encoding for one use of the quantum channel, one also has to take into account the possibility of input states which are entangled across several channel uses. The use of such entangled states can potentially boost the capacity when the channel is used multiple times in parallel as opposed to the use of simple product or separable states. The question of whether such an increase in capacity occurs is commonly referred to as the additivity question. If the use of entanglement can increase the capacity in comparison to product states, we speak of an additivity violation. If, however, such an increase cannot occur, the classical capacity is said to be additive.
In a recent breakthrough, the classical capacity for the so-called phase-insensitive bosonic Gaussian channels has been found, and it was shown that the capacity for these channels is achieved by using products of coherent states for the encoding~\cite{Giovannettietalultimate}. Furthermore, for more general Gaussian channels which are phase-sensitive, the capacity has been found above a certain energy threshold~\cite{Giovannettietalultimate}. This was achieved by proving the optimality of Gaussian inputs above the energy threshold, while the capacity restricted to Gaussian inputs above the energy threshold had been calculated in earlier work~\cite{Pilyavets_2012,Schaefer_2011}. Despite these landmark results, the classical capacity of a general Gaussian channel is not known for all energies, and very little is known about the capacity of non-Gaussian bosonic channels, which are rarely considered in literature.
However, recent developments in the area of entropy power inequalities~\cite{KoeSmiEPI,MariPalma15,HubKoeVersh16} and the minimum output entropies of bosonic channels~\cite{De_Palma_2016passive,De_Palma_2016_maximizers,De_Palma_2017attenuator,De_Palma_2017} make it possible to give bounds on the output entropy of non-Gaussian bosonic channels. These inequalities promise to help at least partially resolve the question of classical capacity for these channels. Here we apply entropy power inequalities to give upper and lower bounds on the classical capacity of a wide class of bosonic noise channels, which includes non-Gaussian channels. 

The channels we consider are attenuator channels with general environment states which are possibly non-Gaussian and classical noise channels with general additive noise.
While we cannot show that the classical capacity of these channels is additive, we can show that additivity violations, if existent, are minor. By this we mean that the maximal difference between the full capacity and the one-shot capacity is a constant independent of the input energy. Furthermore, the lower bounds we obtain are achievable by simple coherent state coding. As mentioned, this coding achieves the capacity for phase-insensitive Gaussian channels. It is, however, known not to be optimal in general even for Gaussian channels~\cite{Lupo_2009}. Nonetheless, as our results show, it still is a suitable choice of encoding for the large class of bosonic channels we consider (which includes the Gaussian channels for which the capacity is known as special cases).

Our paper is structured as follows: In Section \ref{sec:cap}, we review the notion of capacity and related formulas and give a technical introduction to the problem we consider. In Section \ref{sec:bosonicgeneral}, we
state the relevant entropy power and photon number inequalities and give an overview of the techniques we  apply. In the following Section \ref{sec:results}, we state our bounds on the classical capacity of the channels. We present the proofs in Section \ref{sec:proofs}. We close with a discussion of our work and related problems as well as directions for future research.

\section{Capacity of channels}
\label{sec:cap} 

\noindent In the classical setting, where a channel is simply a conditional distribution $P_{Y | X}$ describing the channel's probabilistic output $Y$ on a given input $X$, the classical capacity has been studied by Shannon in his landmark paper~\cite{Shannon48}. It is remarkable that although the definition of the capacity involves  an
arbitrarily large number of channel uses, in the classical setting 
it can be expressed in terms of an optimization problem defined in terms of a single channel use only. It is given by 
\begin{equation}
C\left(P_{Y|X}\right) = \sup_{P_X} I(X:Y)\ ,\label{eq:mutualinfoboundc}
\end{equation}
where $I(X:Y)=H(X)+H(Y)-H(XY)$ is the mutual information between the input and output of the channel, defined in terms of the Shannon entropy $H(X)=-\sum_{x}P_X(x)\log P_X(x)$. In the continuous-variable case of interest here, $X$ and $Y$ are random variables on $\mathbb{R}$, and an energy constraint needs to be imposed on the distributions $P_X$ in~\eqref{eq:mutualinfoboundc}:
the capacity with input ``energy'' $E$ is then defined as in~\eqref{eq:mutualinfoboundc}, but with a constraint $\mathbb{E}[X^2]\leq E$ on the second moment. This constraint amounts to the demand that in the operational coding problem defining the capacity, only codewords, i.e., sequences $m_x=(m_1,\ldots,m_n)\in\mathbb{R}^n$ of elements having a mean energy $\frac{1}{n} \sum_{i=1}^n m_i^2$ bounded by $E$ are allowed. We will denote the corresponding capacity by $C_{\mathrm{E}}(P_{Y|X})$. 

In the quantum setting, a quantum channel is a completely positive trace-preserving (CPTP) map~$\cE:\mathcal{B}(\cH)\rightarrow\mathcal{B}(\cH)$ 
on the set~$\mathcal{B}(\cH)$ of bounded operators on~$\cH$. 
For the bosonic systems considered here, $\cH\cong L^2(\mathbb{R})$ is an
infinite-dimensional separable Hilbert space. We are concerned with the energy-constrained classical capacity 
of such channels: it is operationally defined as the maximal achievable rate~$R$
at which classical bits can be sent by (i)~encoding a message $x\in \{0,1\}^{\lfloor nR\rfloor }$ into a state~$\rho_x$ on $\cH^{\otimes n}$, and (ii)~decoding the received state~$\cE(\rho_x)$ using a suitable POVM~$\{F_x\}_{x\in \{0,1\}^{\lfloor nR\rfloor} }$ on $\cH^{\otimes n}$, in such a way that the average decoding error probability vanishes in the limit~$n\rightarrow\infty$.  In this operational problem, the energy constraint amounts to imposing the physical restriction that the mean photon number $\frac{1}{n}\tr(\sum_{j=1}^n a_j^\dagger a_j\overline{\rho})$ of the average input state $\overline{\rho}$  is bounded by some constant~$N$. Here $a^\dagger_j, a_j$ are the creation and annihiliation operators of the $j$-th mode of a system of $n$ harmonic oscillators, satisfying the canonical commutation relations $[a_j, a_k^\dagger] = \delta_{jk}\id, [a_j, a_k] = 0$.

An expression for the classical capacity of a quantum channel  similar to~\eqref{eq:mutualinfoboundc} is known: the Holevo-Schumacher-Westmoreland theorem (HSW theorem)~\cite{schumacherwestmore,holevocapacity} states that the classical capacity of a quantum channel $\cE$ subject to the energy constraint~$N$ can be obtained
by evaluating the limit
\begin{equation}
  C_N(\cE) = \lim_{n \rightarrow \infty} \frac{1}{n} \chi_{nN}\left(\cE^{\otimes n}\right).
  \label{eq:ccap}
\end{equation}
In this expression, 
$\chi_{nN}(\cE^{\otimes n})$
is the  Holevo quantity
evaluated for the channel $\cE^{\otimes n}:\cB(\cH^{\otimes n})\rightarrow\cB(\cH^{\otimes n})$ with average energy constraint~$N$. The latter is defined as 
\begin{equation}
  \chi_{nN}(\cE^{\otimes n}) = \sup_{\{ p_x, \rho_x\}_x } S\big(\cE^{\otimes n}(\overline{\rho})\big) - \sum_x p_x S\big(\cE^{\otimes n}(\rho_x)\big) \ ,
\label{eq:holevo}
\end{equation}
where  $S(\rho) = - \tr \rho \log \rho$ denotes the von Neumann entropy, and 
where the optimization is over all ensembles 
$\{p_x, \rho_x\}_x$ of states  on $\cH^{\otimes n}$ with average signal state $\overline{\rho} = \sum_x p_x \rho_x$ satisfying the average energy constraint $\tr\left(\sum_{j=1}^n a_j^\dagger a_j \overline{\rho}\right) \leq nN$. In general, one may also consider continuous ensembles of the form $\{p(x)\mathrm{d}x,\rho_x\}_x$ where $dx$ is e.g., the Lebesgue measure on $\mathbb{R}^n$, and $p$ is a probability density function. In this case, sums need to be replaced by integrals.

Expression~\eqref{eq:ccap} for the classical capacity generalizes formula~\eqref{eq:mutualinfoboundc} to quantum channels. Unfortunately, though, it is generally intractable both numerically and analytically: it requires  optimization over an arbitrarily large number~$n$  of copies of the channel. The regularization (i.e., the process of taking the limit $n\rightarrow\infty$ in~\eqref{eq:ccap}) is
necessary to allow for input (code) states which are entangled across several channel uses: such states may potentially improve upon the capacity compared to product states. It is worth noting that the use of separable states as input states, which are neither entangled nor product states, does not improve the capacity in comparison to the use of product states only. This follows from the fact that it is enough to consider pure input states~\cite{schumacherwestmore} and the fact that pure separable states are product states. The capacity when one restricts to codes using only unentangled signal states, that is, the {\em one-shot capacity}, is 
given by~$\chi_N(\cE)$. It gives the lower bound
\begin{align}
C_N(\cE)\geq \chi_N(\cE)\label{eq:boundholevocapacity}
\end{align}
on the classical capacity.

The additivity problem consists in the question of whether or not the inequality~\eqref{eq:boundholevocapacity} is strict (in which case we speak of an additivity violation), or simply an equality. Its name derives from the fact that if one has 
\begin{align}
\chi_{nN}(\cE^{\otimes n})=n\chi_N(\cE)\qquad\textrm{ for all } n\in\mathbb{N}\ ,\label{eq:holevoquantityadditivity}
\end{align}
then one immediately obtains equality in~\eqref{eq:boundholevocapacity}, implying that entangled signal states offer no operational advantage.

While it is still unknown whether or not equality holds in~\eqref{eq:boundholevocapacity} in general in the continuous-variable case, the simpler additivity property~\eqref{eq:holevoquantityadditivity} for the Holevo quantity has been shown not to hold in general by Hastings~\cite{hastings}. He  showed that there exists a channel $\cT:\cB(\mathbb{C}^d)\rightarrow\cB(\mathbb{C}^d)$ for which
\begin{equation}
  \chi(\cT^{\otimes 2}) > 2 \chi(\cT)\ .
\end{equation}
(There is no energy constraint here because only finite-dimensional Hilbert spaces are involved.) 
In principle, this leaves room for an improvement upon the classical capacity via the use of entangled signal states, i.e., the inequality~\eqref{eq:boundholevocapacity} may still be strict for certain channels. Understanding when this may or may not be the case is one of the central challenges of quantum information theory, and fits into the larger theme of investigating the impact of quantum effects such as entanglement  on the power of information-processing primitives. \\

\subsection*{Bosonic noise channels}
Here we explore the potential of additivity violations in channels associated with bosonic systems. The quantum channels discussed here are attenuation channels (i.e. beamsplitters coupling the system to an environment in a general state) as well as channels mixing the system with a classical random variable. These are natural generalizations of the corresponding Gaussian channels: the thermal noise channel and the classical noise channel.
These Gaussian channels have been the subject of various earlier analyses~\cite{Giovannettietalultimate,HolevoWernerGaussian,KoeSmiEPIchannel,KoeSmi12b,Giovannetti_2013},
and, as discussed below, have been shown not to violate additivity in a recent breakthrough development. In contrast, our emphasis here is on general, potentially non-Gaussian bosonic channels, for which no additivity statements have been known previously.

In more detail, we consider an input system of $d$ bosonic modes (with Hilbert space $\cH^{\otimes d}$ where $\cH\cong L^2(\mathbb{R})$)  with the vector of mode operators $R = (Q_1,P_1, \dots, Q_d, P_d)$.
The action of a beamsplitter with transmissivity $0 \leq  \lambda \leq  1$ which couples the system with an environment of $d$ bosonic modes with mode operators $(Q^{\mathrm{(E)}}_1, P^{\mathrm{(E)}}_1, \dots, Q^{\mathrm{(E)}}_d, P^{\mathrm{(E)}}_d)$ 
is given by a Gaussian unitary~$U_\lambda$ on $\cH^{\otimes d}\otimes\cH^{\otimes d}$. Its  action on the $2d$ modes is defined (in the Heisenberg picture)  by the symplectic matrix
\begin{equation}
  S_{\lambda} = \begin{pmatrix}\sqrt{\lambda}\id_{2d} & \sqrt{1-\lambda}\id_{2d} \\ \sqrt{1-\lambda}\id_{2d} & -\sqrt{\lambda}\id_{2d} \end{pmatrix}
\end{equation}
with respect to the ordering $(Q_1,P_1,\dots, Q_d, P_d, Q^{\mathrm{(E)}}_1, P^{\mathrm{(E)}}_1, \dots, Q^{\mathrm{(E)}}_d, P^{\mathrm{(E)}}_d)$ of modes, i.e., $U_\lambda^\dagger R_j U_\lambda =\sum_{k}(S_\lambda)_{j,k} R_k$. If we assume that the environment is in some state~$\sigma_{\mathrm{E}}$ (decoupled from the system), and consider only the action of this unitary on the system, we obtain the quantum channel
\begin{equation}
  \cE_{\lambda, \sigma_{\mathrm{E}}}(\rho) := \tr_{\mathrm{E}} \left(U_\lambda (\rho \otimes \sigma_{\mathrm{E}}) U_\lambda^\dagger \right)\ .
  \label{eq:beamsplit}
\end{equation}
We are interested in the classical capacity of channels~$\cE_{\lambda,\sigma_{\mathrm{E}}}$ of the form~\eqref{eq:beamsplit}, which we call attenuation channels. Note that this set of channels includes Gaussian channels (for Gaussian states~$\sigma_{\mathrm{E}}$) such as the thermal noise channels (when $\sigma_{\mathrm{E}}=e^{-\beta (\sum_{j=1}^d Q_j^2+P_j^2)}/Z$ is a thermal state, i.e., the Gibbs state of a certain quadratic Hamiltonian) or the pure loss channel (when $\sigma_{\mathrm{E}}$ is the vacuum state). It also includes non-Gaussian channels (for $\sigma_{\mathrm{E}}$ a non-Gaussian state): typical examples include, e.g., the case where
$\sigma_{\mathrm{E}}$ slightly deviates from a thermal state, or is, e.g., some finite superposition of number states.

A second class of channels we consider here 
are channels which act by displacing the system
according to some 
probability density function~$f:\mathbb{R}^{2d}\rightarrow\mathbb{R}$ on phase space. We call these (general) classical noise channels. They act as
\begin{equation}
  \cF_{t,f}(\rho) := \int f(\xi)W(\sqrt{t}\xi) \rho W(\sqrt{t}\xi)^\dagger \mathrm{d}^{2d}\xi\ ,
\label{eq:cqchannel}
\end{equation}
where $W(\xi) = e^{i\sqrt{2\pi} \xi \cdot (\sigma R)}$ for $\xi \in \mathbb{R}^{2d}$ are the Weyl displacement operators with the symplectic form $\sigma = \begin{pmatrix}0 & 1 \\ -1 & 0\end{pmatrix}^{\oplus d}$ and the mode operators~$R~=~(Q_1, P_1, \dots, Q_d, P_d)$. Here $t > 0$ is some parameter analogous to the transmissivity. Again, this channel may be non-Gaussian (if $f$ is not a Gaussian distribution). Channels of this type have been considered by Werner~\cite{WernerHarmonicanalysis84}, who described them as a convolution operation between a probability distribution and a state. They satisfy a number of convenient properties with respect to displacements in phase space as well as a data processing inequality. For more background we refer to \cite{HubKoeVersh16,WernerHarmonicanalysis84,dattaetal16}.

For a specific kind of quantum channels, the so-called single-mode phase-insensitive\footnote{

  A phase-insensitive channel is a channel $\Phi$ which has one of the following properties under phase shift operations $e^{i\varphi a^\dagger a}$~\cite{Giovannettietal04,Holevo_2015}:
  $\Phi(e^{i\varphi a^\dagger a} \rho e^{-i\varphi a^\dagger a}) = e^{i\varphi a^\dagger a} \Phi(\rho) e^{-i\varphi a^\dagger a}$ for all $\rho$ for gauge-covariant channels and with reversed order of operators on the rhs. for gauge-contravariant channels.
  }
Gaussian channels, the classical capacity has recently been found by Giovannetti \etal~\cite{Giovannettietalultimate,Giovannetti_2014}. 

The channels~\eqref{eq:beamsplit},~\eqref{eq:cqchannel} fall into this class if the environment state $\sigma_{\mathrm{E}}$ is a thermal state or if the probability density function~$f$ is a Gaussian distribution whose covariance matrix is proportional to the identity (this special case has commonly been referred to as the classical noise channel, see~\cite{HolevoWernerGaussian,eisertwolfb}). In particular, if $\sigma_{\mathrm{E}}$ is a single-mode Gaussian thermal state, the capacity of the single-mode  ($d=1$) channel $\cE_{\lambda,\sigma_{\mathrm{E}}}$ is given by
\begin{equation}
  C_N(\cE_{\lambda, \sigma_{\mathrm{E}}}) = g\big(\lambda N + (1-\lambda) N_{\mathrm{E}}\big) - g\big((1-\lambda)N_{\mathrm{E}}\big)\ ,
\label{eq:gausscap}
\end{equation}
with the mean photon number $N_{\mathrm{E}} = \tr(a^\dagger a \sigma_{\mathrm{E}})$ of the environment. Furthermore, if $f$ is a centered Gaussian distribution of unit variance, then the single-mode channel $\cF_{t,f}$ has capacity~\cite{Giovannettietalultimate}
\begin{equation}
C_N(\cF_{t, f}) = g(N+2\pi t) - g(2\pi t)\ .
\label{eq:cqcap}
\end{equation}
In both cases, the quantities on the rhs.~of Eq.~\eqref{eq:gausscap}
and Eq.~\eqref{eq:cqcap} have been shown to be equal to the single-shot
Holevo information ($\chi_N(\cE_{\lambda,\sigma_{\mathrm{E}}})$ and $\chi_N(\cF_{t,f})$, respectively). In particular, there is no violation of additivity in these channels and thus no operational gain in using entangled signal states. 
Similar capacity formulas have been found for single-mode phase-sensitive Gaussian channels, where the environment is in a Gaussian state $\sigma_\mathrm{E}$ which might be squeezed (and respectively, if the function $f$ is a Gaussian whose covariance matrix is not proportional to the identity). In this case the capacity is only known if the energy constraint allows for input energies larger than a certain threshold value~\cite{Giovannettietalultimate,Pilyavets_2012,Schaefer_2011}. 

This fundamental result is striking, but leaves open the question of whether
additivity violation is possible in more general channels. 
This is one motivation for considering the more general families~$\{\cE_{\lambda,\sigma_{\mathrm{E}}}\}$ and $\{\cF_{t,f}\}$ of single-mode channels, which also include non-Gaussian examples.

We find that additivity violations, if at all existent, must be limited: 
the difference $C_N(\Phi)-\chi_N(\Phi)$ between the two sides of~\eqref{eq:boundholevocapacity} is upper bounded by a constant independent of the input photon number $N$, for any channel~$\Phi$ in the class of attenuators and classical noise channels. In other words, we show that the maximal potential gain achievable by entangled coding strategies is limited. This means that for these channels, the use of entanglement cannot improve the classical capacity achieved by classical modulation of coherent states by much: with growing input energies, the maximal gain by coding strategies using entanglement becomes negligible compared to the value of the capacity. 

Our work follows similar reasoning as that of K\"onig and Smith~\cite{KoeSmiEPIchannel}, who addressed the  question whether entangled coding strategies can substantially increase the classical capacity of thermal noise channels: we also employ entropy power inequalities to obtain upper bounds on the capacity. However, in contrast to~\cite{KoeSmiEPIchannel}, we also need to establish new achievability (lower) bounds on the capacity: here we again use entropy power inequalities, as well as Gaussian extremality -- this reasoning follows pioneering work by Shannon~\cite{Shannon48}. While the results of~\cite{KoeSmiEPIchannel} for the thermal noise channel have by now been superseded by the explicit capacity formulas of~\cite{Giovannettietalultimate}, 
it appears unlikely that similar explicit formulas can be established
with present-day analytical methods for the non-Gaussian channels discussed here.

\section{Analytical tools for bosonic systems\label{sec:bosonicgeneral}}
Evaluating the classical capacity amounts to solving a highly non-trivial optimization problem. Even for the one-shot capacity, which does not
involve an  infinite limit over parallel uses of the channel,
explicit capacity formulas are generally not known. In special cases, e.g., when the channel exhibits certain symmetries (in particular, gauge-covariance or contravariance in the bosonic context), this difficulty can be overcome~\cite{Giovannettietalultimate,Garcia_Patron_2014,Holevo_2016}. However, the focus of our work is on more general, possibly non-Gaussian channels. In this context, only few analytical tools are known: these include entropy power inequalities (EPI), the Gaussian maximum entropy principle, as well as certain more recent Gaussian optimizer results. In this section, we briefly review these results, and also discuss related conjectures. In Section~\ref{sec:results}, we then discuss the implications
of these statements to classical capacities: we will see that
they imply various bounds
on the possible degree of non-additivity.

\subsection{Gaussian extremality}
A key tool in dealing with non-Gaussian distributions, states and optimization are Gaussian extremality results. The main result we use here
is Gaussian extremality for the von Neumann entropy (but see~\cite{Wolfetal06} for more general statements and applications): this states that among all states with fixed first and second moments, the 
Gaussian state has maximal entropy. Succinctly, this can be expressed by the inequality
\begin{align}
  S(\rho) \leq S([\rho])\ ,\label{eq:gaussianextremalityinequality}
\end{align}
where $[\rho]$ is the Gaussian state with the same first and second moments as $\rho$.  As a corollary, among all one-mode states~$\rho$ with mean photon number~$\tr(a^\dagger a\rho)$ smaller or equal to $N$, the Gaussian thermal state 
$\rho_{\mathrm{th},N} = \frac{1}{N+1}\sum_{n=0}^\infty \left(\frac{N}{N+1}\right)^n\proj{n}$ has maximal entropy $g(N):=~(N+1)\log(N+1)-N\log(N)$. A simple proof of this corollary can be found, for instance, in \cite[Lemma 9]{depalmaconditional}, while the entropy of Gaussian states has been calculated in~\cite{Agarwal_1971}.

Gaussian states also turn out to be optimal for various entropy or entropy-related quantities defined in terms of Gaussian operations.  For example, Gaussian states have recently been shown to be the optimizers of the defining problem of $\|\Phi\|_{p\rightarrow p}$-norms for a Gaussian channel~$\Phi$, see~\cite{Frank_2017,Holevopq}
In a similar context, in a series of recent works by De Palma, Trevisan, and Giovannetti \cite{De_Palma_2016passive,De_Palma_2016_maximizers,De_Palma_2017attenuator, De_Palma_2017}, it was  shown that the output entropy of any gauge-covariant one-mode Gaussian channel for a fixed input entropy is minimized by taking as input state the thermal state of this fixed input entropy. For the beamsplitter with environment in the thermal state~$\rho_{\mathrm{th},N}$, this can be stated as \cite[Theorem 4]{De_Palma_2017}
\begin{equation} 
  S(\cE_{\lambda,\rho_{\mathrm{th}}}(\sigma)) \geq g\big(\lambda g^{-1}[S(\sigma)] + (1-\lambda) N\big)\ .
  \label{eq:minout}
\end{equation}
As explained below, Eq. \eqref{eq:minout} is a special case of the currently unproven entropy photon number inequality (EPNI) conjecture, which is of relevance to this work.

\subsection{Entropy power inequalities}
\label{sec:setup}
In classical information theory, the Shannon entropy of  an $\mathbb{R}^n$-valued  random variable $X$ with probability density $f:\mathbb{R}^n\rightarrow\mathbb{R}$ is given by $H(X) = - \int f(x) \log f(x) \mathrm{d}^nx$. In order to estimate the capacity of additive noise channels, Shannon~\cite{Shannon48-2} proposed the entropy power inequality (EPI)
\begin{equation}
  e^{2H(X+Y)/n} \geq e^{2H(X)/n} + e^{2H(Y)/n}\ ,
  \label{eq:epi}
\end{equation}
where the lhs. is the entropy power of the sum of two independent random variables $X$ and~$Y$. A rigorous proof of~\eqref{eq:epi} was established by Stam~\cite{Stam59,Blachman65} 
under the assumption that $X$ and $Y$ are of finite variance. Blachman in~\cite{Blachman65} gave a detailed account of Stam's proof, and Lieb in~\cite{Lieb78} subsequently found a different proof of the entropy power inequality using Young's inequality for convolutions.

In the context of quantum information theory, Shannon's entropy power inequality has been generalized. 
In~\cite{KoeSmiEPI}, the inequality
\begin{equation}
  e^{S\big(\cE_{\lambda,\sigma_{\mathrm{E}}}(\rho)\big)/n} \geq \lambda e^{S(\rho)/n} + (1-\lambda) e^{S(\sigma_{\mathrm{E}})/n}\ 
    \label{eq:qepi}
  \end{equation}
was shown for $\lambda=1/2$. This was then generalized by De Palma et al.~\cite{MariPalma15} to all $\lambda\in [0,1]$. 
The lhs. involves the von Neumann entropy of the output~$\cE_{\lambda,\sigma_{\mathrm{E}}}(\rho)$ under a beamsplitter of transmissivity $\lambda$, defined in~\eqref{eq:beamsplit},
and can be considered as a quantum convolution operation of two $n$-mode states $\rho$ and $\sigma_{\mathrm{E}}$, analogous to the convolution $X+Y$ of two independent random variables~$X$ and $Y$.  More recently, a conditional version of the entropy power inequality has been proven by De Palma and Trevisan~\cite{depalmaconditional}, generalizing a statement previously established for  Gaussian states only~\cite{koenigconditionalepi2015}. This result can be stated as
\begin{align}
  \exp \frac{S(C|E)_{\rho_{CE}}}{n} \geq \lambda \exp \frac{S(A|E)_{\rho_{AE}}}{n} + (1-\lambda) \exp \frac{S(B|E)_{\rho_{BE}}}{n}\label{eq:conditionalepi}
\end{align}
for the input system $A$, the environment system $B$, and the output system $C$ of the beamsplitter.  This concerns
a tripartite state~$\rho_{ABE}$ consisting of two $n$-mode systems $A$ and $B$  which are conditionally independent given $E$,
as well as  the result~$\rho_{CE}=\tr_B \left((U_{\lambda}\otimes I_E)\rho_{ABE}(U_{\lambda}\otimes I_E)^\dagger\right)$ of applying a transmissivity-$\lambda$ beamsplitter~$U_\lambda$ to $AB$.
Remarkably,
the proof presented in~\cite{depalmaconditional} circumvents all regularity assumptions required in earlier proofs: it is valid for any state $\rho_{ABM}$ with finite mean photon number (second moments) in~$AB$ and  satisfying $S(\rho_E) < \infty$.
This also implies the validity of~\eqref{eq:qepi} for all
states $\rho$ and $\sigma_{\mathrm{E}}$ with finite mean photon number. The conditional entropy power inequality~\eqref{eq:conditionalepi} has application to establishing upper bounds on the entanglement-assisted classical capacity of bosonic quantum channels, as proposed in~\cite{koenigconditionalepi2015}.

Another generalization of \eqref{eq:epi} has been established in \cite{HubKoeVersh16} and can be stated as
\begin{equation}
  e^{S\big(\cF_{t,f}(\rho)\big)/n} \geq e^{S(\rho)/n} + t e^{H(f)/n}\ ,
  \label{eq:cqepi}
\end{equation}
for a probability density function $f:\mathbb{R}^{2n}\rightarrow\mathbb{R}$ and an $n$-mode state $\rho$. 
The lhs.~of this inequality involves the entropy power of the output~$\cF_{t,f}(\rho)$ of the classical noise channel, defined in~\eqref{eq:cqchannel}. As before, the expression~$\cF_{t,f}(\rho)$ can be considered as a convolution operation, in this case between a probability density function~$f$ and a quantum state~$\rho$. 
The proof presented in~\cite{HubKoeVersh16} follows earlier heat-flow arguments and requires certain regularity assumptions. It appears straightforward, however, to adapt the proof of~\cite{depalmaconditional} to this setting -- 
this would show that~\eqref{eq:cqepi} holds for all probability density functions $f$ with finite second moments and all states~$\rho$ with finite mean photon number. 

\subsection{Conjectured entropy photon number inequalities}

In \cite{Guhaetal07} and \cite{photonnumbersecond}, an alternative to the entropy power inequality \eqref{eq:qepi} has been proposed, replacing the entropy power of $\rho$ by the mean photon number of a Gaussian thermal state with the same entropy. This inequality is called the Entropy Photon-Number Inequality (EPNI) and can be stated as
\begin{equation}
  g^{-1}\left(\frac{S(\cE_{\lambda,\sigma_{\mathrm{E}}}(\rho))}{n} \right) \geq \lambda g^{-1}\left(\frac{S(\rho)}{n}\right) + (1-\lambda)g^{-1}\left(\frac{S(\sigma_{\mathrm{E}})}{n}\right)\ ,
\label{eq:epni}
\end{equation}
where the channel $\cE_{\lambda,\sigma_{\mathrm{E}}}$ is used in parallel on $n$ modes. This statement can be seen as a generalization of \eqref{eq:minout} to multiple modes and the case when the environment is not a thermal state. 
Despite progress in certain special cases \cite{Das_2013}, and the special case of \eqref{eq:minout}, the EPNI remains unproven. However, its implications on capacities have been studied in a number of works \cite{Guhaetal07,photonnumbersecond,SaikatGuhaetal07}. 

It is natural to ask whether an EPNI also holds in the case of the channel $\cF_{t,f}$. We conjecture that this is the case and that the ``classical-quantum'' EPNI reads
\begin{equation}
  g^{-1}\left(\frac{S(\cF_{t,f}(\rho))}{n}\right) \geq g^{-1}\left(\frac{S(\rho)}{n}\right) + \frac{t}{e} e^{\frac{H(f)}{n}}.
  \label{eq:cqepni}
\end{equation}
A weaker statement which would still be useful for establishing lower bounds on the classical capacity is the specialization of this EPNI to the case of one mode and a thermal input state:
\begin{equation}
  g^{-1}\big[S\big(\cF_{t,f}(\rho_{\mathrm{th},N})\big)\big] \geq N + \frac{t}{e} e^{H(f)}\ ,
  \label{eq:minoutconj}
\end{equation}
where $\rho_{\mathrm{th},N}$ is the Gaussian thermal state with mean photon number $N$ as introduced above. This inequality can be seen as a classical noise channel analog of Eq.~\eqref{eq:minout} for the attenuator. A proof of Eq. \eqref{eq:minoutconj} might be easier than the full proof of Eq. \eqref{eq:cqepni} and still have desirable implications: we show that assuming the validity of Eqs.~\eqref{eq:minoutconj} and \eqref{eq:cqepni} leads to better bounds than using the entropy power inequality, again without altering the spirit of the theorems.

Having reviewed the key tools required for the discussion, we continue with the statement of our results in the next section.

\section{Limited  non-additivity for non-Gaussian bosonic channels}
\label{sec:results}
Here we show that the recent generalizations of the EPI as well as 
the related results discussed in Section~\ref{sec:bosonicgeneral}
have direct implications for the classical capacities of non-Gaussian bosonic channels. In particular, they imply that the degree of potential non-additivity is limited for
attenuators and classical noise channels. A similar analysis has been carried out in~\cite{KoeSmiEPIchannel} for special cases of the channels considered here, namely Gaussian thermal noise channels. In contrast to that work, we show that the degree of non-additivity can also be bounded for non-Gaussian channels. This extends our understanding of classical capacities to previously untreatable cases.  

One of the key observation is that EPIs can be used to obtain 
not only upper (converse) bounds on the classical capacity of non-Gaussian channels, but also achievability bounds. Remarkably, these achievability bounds concern simple product state codes consisting of coherent states. This coding strategy has recently been shown to be optimal for phase-insensitive Gaussian channels. It is known to not be optimal~\cite{Giovannettietalultimate,Pilyavets_2012,Schaefer_2011} for certain phase-sensitive Gaussian channels, where coding with squeezed coherent states achieves a higher rate. 
Nonetheless, coherent state coding gives a good lower bound also in these cases, as we will see below.
 
Our work implies that the same strategy of using coherent states is also essentially optimal for the more general non-Gaussian channels considered here.
Thus although this coding strategy is in general not optimal even for Gaussian channels, we show that it is a suitable choice of coding strategy for non-Gaussian channels. 

In spirit, our results can be seen as quantum generalizations of Shannon's work, in which he applied the entropy power inequality to the capacity of the additive noise channel~\cite{Shannon48-2}: He found that the capacity $C_P$ of a classical additivity channel $Y=X+Z$, 
where $Z$ is noise independent of the input $X$ (but otherwise arbitrary), is bounded by
\begin{equation}
\log \frac{P + N_1}{N_1} \leq C_P \leq \log \frac{P + N}{N_1}\ , \label{eq:shannonbounds}
\end{equation}
where $P$ is the average transmitter power, $N$ is the average noise power, and $N_1 = e^{2H(Z)}/(2\pi e)$ is the entropy power of the noise $Z$. In the special case where $Z$ is distributed according to
a standard normal distribution, the upper and lower bounds in Eq.~\eqref{eq:shannonbounds} coincide and reduce to  Shannon's capacity formula for the additive white Gaussian noise channel. 

Our main result concerns the single-mode attenuation channel $\cE_{\lambda,\sigma_{\mathrm{E}}}$ introduced in~\eqref{eq:beamsplit},
and the classical noise channel $\cF_{t,f}$ introduced in~\eqref{eq:cqchannel}.  We shall denote
the mean photon number~$\tr(a_{\mathrm{E}}^\dagger a_{\mathrm{E}} \sigma_{\mathrm{E}})$ of $\sigma_{\mathrm{E}}$ by $N_{\mathrm{E}}$, and its von Neumann entropy by $S(\sigma_{\mathrm{E}})$. Similarly, we write 

\begin{equation}
  \mathrm{E}(f) = \sum_{i=1}^2 \int \xi_i^2 f(\xi) \mathrm{d}^2\xi
\end{equation}

for the energy (i.e., the sum of second moments) of the distribution $f$ and

\begin{equation}
  H(f)=-\int f(\xi)\log f(\xi)\mathrm{d}^2\xi
\end{equation}

for the Shannon entropy of the associated random variable. Throughout, we will assume that these quantities are finite. We then have the following main result:

\begin{theorem}\label{thm:main}
  The maximal degree of non-additivity of the classical capacity of $\cE_{\lambda,\sigma_{\mathrm{E}}}$ and~$\cF_{t,f}$  is bounded as
\begin{align}\noindent
  C_N(\cE_{\lambda,\sigma_{\mathrm{E}}})-\chi_N(\cE_{\lambda,\sigma_{\mathrm{E}}})& \leq 
  2 g\big( (1-\lambda)N_{\mathrm{E}}\big) - g\big( (1-\lambda) N_{\mathrm{E}}^{\mathrm{ep}}\big) - \log\left(\lambda + (1-\lambda) e^{S(\sigma_{\mathrm{E}})}\right)\ ,\\
C_N(\cF_{t,f})-\chi_N(\cF_{t,f})&\leq 2g(\pi t \mathrm{E}(f)) - \log\left( 1 + t e^{H(f)}\right)\ ,
\end{align}
independently of the  input energy $N$, where $N_{\mathrm{E}}^{\mathrm{ep}} = g^{-1}\left[S(\sigma_{\mathrm{E}}) \right]$ is the mean photon number of a thermal state with the same entropy as $\sigma_\mathrm{E}$.
For both channels,  coherent states modulated  with a Gaussian distribution achieve a rate which differs from the capacity by at most the rhs.~of these bounds.
\end{theorem}
The fact that the rhs.~of these bounds is independent of the input energy $N$  implies  that the degree of violation is at most constant. 
In particular, the potential violation is
negligible compared to the actual value of the capacity for large~$N$. In other words, there is no significant advantage in using entangled states for coding. This result is indeed not surprising: For very large values of $N$, the associated quantum channels are ``almost classical'' and therefore it is natural to expect quantum effects such as non-additivity to become small in this regime. 

The maximal degree of violation depends on the structure of the environment (the state $\sigma_{\mathrm{E}}$ respectively the distribution $f$) and is not simply a universal constant as in~\cite{KoeSmiEPIchannel}. This is not surprising because~\cite{KoeSmiEPIchannel} only considered attenuation channels with Gaussian thermal states in the environment (also called thermal noise channels).

Unlike Shannon's result~\eqref{eq:shannonbounds}, the bounds in Theorem~\ref{thm:main} do not specialize to
the known capacity results for Gaussian channels in the case where  $\sigma_{\mathrm{E}}$ is a thermal state or $f$ is a unit-variance centered normal distribution. We show that stronger bounds with this property can be derived assuming the  validity of the EPNI conjecture:

\begin{theorem} \label{thm:bound_additive}
  Assuming the EPNI conjecture \eqref{eq:epni} holds, we have 
\begin{align}
  C_N(\cE_{\lambda,\sigma_{\mathrm{E}}}) -\chi_N(\cE_{\lambda,\sigma_{\mathrm{E}}})&\leq 2\bigg[g\big( (1-\lambda) N_{\mathrm{E}}\big) - g\big( (1-\lambda) N_\mathrm{E}^\mathrm{ep}\big) \bigg]\ .
\end{align}
Similarly, assuming that  the EPNI conjecture \eqref{eq:cqepni} holds, we have
\begin{align}
  C_N(\cF_{t,f}) -\chi_N(\cF_{t,f})&\leq 2\bigg[g(\pi t \mathrm{E}(f)) - g\left(\frac{t}{e}e^{H(f)}\right) \bigg] \  .
\end{align}
Furthermore, coherent state modulation with a Gaussian distribution achieves a rate which differs from the capacity by at most the rhs.~of these bounds.
\end{theorem}
We  stress that these bounds hold independently of whether or not $\sigma_{\mathrm{E}}$ or $f$ (and thus the channels) are Gaussian. In the special case where $\sigma_{\mathrm{E}}$ is a thermal state, we have $N_\mathrm{E}^\mathrm{ep} = N_{\mathrm{E}}$: here Theorem~\ref{thm:bound_additive}
implies that there is no additivity violation, and  we recover the Gaussian capacity result for the thermal noise channel~\eqref{eq:gausscap} (see below).  Similarly, if $f$ is a unit-variance centered normal distribution, Theorem~\ref{thm:bound_additive} reduces to the capacity result~\eqref{eq:cqcap} for the Gaussian classical noise channel. 
In this respect,  Theorem~\ref{thm:bound_additive} behaves similarly as Shannon's bounds~\eqref{eq:shannonbounds} and is compatible with the capacity formulas of~\cite{Giovannettietalultimate}.

  Furthermore, if $\sigma_\mathrm{E}$ is a squeezed thermal state, coherent state coding is known not to be optimal~\cite{Pilyavets_2012}. The classical capacity for input energies higher than a certain threshold value in this case is achieved by input states which are squeezed coherent states. In this case the rhs. in Theorem~\ref{thm:bound_additive} becomes a bound on the difference in rate between squeezed coherent state coding and coherent state coding. The same holds true in the case when $f$ is a Gaussian whose covariance matrix is not proportional to the identity~\cite{Schaefer_2011}. In these particular cases (above the threshold energy), the gap between the lower and upper bounds we obtain is not due to non-additivity, but simply due to the fact that coherent state coding is not optimal in the one-shot case for these channels. We stress that this does not weaken the statement of our theorems in the case of non-Gaussian channels, which are our main focus and for which our bounds are new.

We point out, however, that Theorem~\ref{thm:main} (which does not require the EPNI conjectures) essentially gives the same qualitative conclusions for non-Gaussian channels, and Theorem~\ref{thm:bound_additive} does not provide additional information. Indeed, consider for example the case of the attenuation channel, with $\sigma_{\mathrm{E}}=\proj{N_{\mathrm{E}}}$ equal to one of the number states~$\ket{N_{\mathrm{E}}}$.  Then both Theorem~\ref{thm:main} and Theorem~\ref{thm:bound_additive} specialize to 
\begin{align}
  C_N(\cE_{\lambda,\proj{N_{\mathrm{E}}}})-
  \chi_N(\cE_{\lambda,\proj{N_{\mathrm{E}}}}) &\leq 2 g\big((1-\lambda)N_{\mathrm{E}}\big)\ .
\end{align}
Observe also that for $N_{\mathrm{E}}=0$, the rhs.~vanishes, showing that
the so-called pure loss channel (with $\sigma_{\mathrm{E}}=\proj{0}$ equal to the vacuum state) does not violate additivity. In particular, this provides a new rederivation of the capacity result of~\cite{purelosscapacity} based on the EPI only.

We present the derivation of these results in Section~\ref{sec:proofs}.

\begin{figure}[th]\centering
  \includegraphics[scale=0.5]{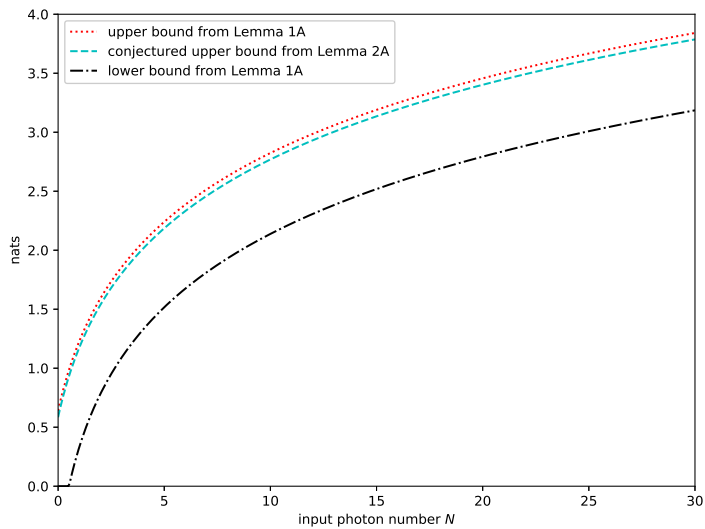}
  \caption{Bounds on the capacity $C_N(\cE_{\lambda,\sigma_{\mathrm{E}}})$ with $\lambda = \frac34$ for any environment state $\sigma_{\mathrm{E}}$ with mean photon number $N_{\mathrm{E}} = 2$ and entropy $S(\sigma_{\mathrm{E}}) \simeq 0.91\; \mathrm{nats} < g(N_{\mathrm{E}}) \simeq 1.91\; \mathrm{nats}$.} 
    \label{fig:1}
\end{figure}

\section{Derivation of non-additivity bounds}
\label{sec:proofs}
In this section, we present the proof of  Theorems~\ref{thm:main}
and~\ref{thm:bound_additive}. Let us first sketch the basic idea: The Holevo quantity~\eqref{eq:holevo} consists of a difference between two entropic quantities. The maximum entropy principle~\eqref{eq:gaussianextremalityinequality} allows us to restrict to Gaussian states when we require upper bounds on entropies (subject to fixed second moments), whereas the entropy power inequalities~\eqref{eq:qepi} and~\eqref{eq:cqepi} (respectively the entropy photon-number inequalities~\eqref{eq:epni} and \eqref{eq:cqepni}), as well as Eq.~\eqref{eq:minout} are suitable to obtain lower bounds on entropies. The bounds on the capacity are then expressions which depend on output entropies restricted to Gaussian input states and Gaussian environments. These quantities can then be bounded with elementary calculations. The derivation of upper bounds on the capacity thus proceeds similarly as in~\cite{KoeSmiEPIchannel}; in addition, we obtain lower bounds on the capacity in a similar fashion.

\subsection{Derivation of capacity bounds from EPIs}
We first consider the attenuation channel~$\cE_{\lambda,\sigma_{\mathrm{E}}}$ introduced in Eq.~\eqref{eq:beamsplit}. The corresponding inequality in Theorem~\ref{thm:main} follows immediately from the following lemma. The corresponding statement for the classical noise channel~$\cF_{t,f}$ is shown in Lemma~\ref{lem:bound_additive1} below.

\subsection{Derivation of bounds from EPIs}
\begin{customlemma}{1A}\label{lem:beamsplitterunconditional}
  The classical capacity of the single-mode attenuation channel $\cE_{\lambda,\sigma_{\mathrm{E}}}$ satisfies
\begin{align}
  C_N(\cE_{\lambda,\sigma_{\mathrm{E}}}) &\geq g\big(\lambda N + (1-\lambda) N_\mathrm{E}^\mathrm{ep}\big) - g((1-\lambda)N_{\mathrm{E}})\ , \label{eq:lowbound1}\\
  C_N(\cE_{\lambda,\sigma_{\mathrm{E}}}) &\leq g\big(\lambda N + (1-\lambda) N_{\mathrm{E}}\big) - \log\left(\lambda + (1-\lambda)e^{S(\sigma_{\mathrm{E}})}\right)\ .\label{eq:upbound1}
\end{align}
The lower bound~\eqref{eq:lowbound1} is achievable with a coherent state ensemble. 
The difference between this upper and lower bound is bounded by 
\begin{equation}
  \Delta(\cE_{\lambda,\sigma_{\mathrm{E}}}) \leq 2 g\big( (1-\lambda)N_{\mathrm{E}}\big) - g\big( (1-\lambda)N_\mathrm{E}^\mathrm{ep}\big) - \log\left(\lambda + (1-\lambda) e^{S(\sigma_{\mathrm{E}})}\right)\ ,\label{eq:differenceboundone}
\end{equation}
independently of the input photon number $N$.
\end{customlemma}
The corresponding upper and lower bounds on the capacity are visualized in Fig.~\ref{fig:1}.

\begin{proof}[Proof of Lemma~\ref{lem:beamsplitterunconditional}]
  \textbf{The upper bound.} We prove the upper bound~\eqref{eq:upbound1} in a similar fashion as~\cite{KoeSmiEPIchannel}. By bounding the Holevo quantity in the Holevo-Schumacher-Westmoreland formula~\eqref{eq:holevo} for the classical capacity  we have
  \begin{equation}
    C_N(\cE_{\lambda,\sigma_{\mathrm{E}}}) \leq S_N^{\mathrm{max}}(\cE_{\lambda,\sigma_{\mathrm{E}}}) - \lim_{n\rightarrow \infty} \frac{1}{n} S^{\mathrm{min}}(\cE_{\lambda,\sigma_{\mathrm{E}}}^{\otimes n})\ ,
    \label{eq:upboundholevo}
  \end{equation}
where
\begin{equation}
  S_N^{\mathrm{max}}(\cE_{\lambda,\sigma_{\mathrm{E}}}) = \sup_{\tr(a^\dagger a\rho) \leq N} S(\cE_{\lambda,\sigma_{\mathrm{E}}}(\rho))
\end{equation}
and  $S^{\min}(\cE) = \inf_{\rho} S(\cE(\rho))$ is the minimum output entropy. For any $n$-mode state $\rho_n$, by the entropy power inequality we have that
\begin{align}
  \frac{1}{n} S\left(\cE_{\lambda,\sigma_{\mathrm{E}}}^{\otimes n} (\rho_n) \right) &\geq \log\left(\lambda e^{S(\rho_n)/n} + (1-\lambda) e^{S\left(\sigma_{\mathrm{E}}^{\otimes n}\right)/n} \right)\\
  &\geq \log\left(\lambda + (1-\lambda) e^{S(\sigma_{\mathrm{E}})}\right) \label{eq:minout1}
\end{align}
by the entropy power inequality~\eqref{eq:qepi} since $S(\rho) \geq 0$ for any state~$\rho$. This is a useful bound on the second term in Eq. \eqref{eq:upboundholevo}.
In order to find a bound on the first term, let us consider the mean photon number of the output state $\cE_{\lambda, \sigma_{\mathrm{E}}}(\rho)$, where the input~$\rho$ has mean photon number bounded by~$N$.  It can be bounded as 
\begin{align}
  \tr\left(a^\dagger a \cE_{\lambda,\sigma_{\mathrm{E}}}(\rho)\right) &=  \lambda \tr\left(a^\dagger a \rho \right) + (1-\lambda) \tr\left(a_{\mathrm{E}}^\dagger a_{\mathrm{E}} \sigma_{\mathrm{E}}\right)\\
  &\leq \lambda N + (1-\lambda) N_{\mathrm{E}}\ . \label{eq:outmax}
\end{align}
Thus the output entropy is bounded as 
\begin{equation}
  S_N^{\mathrm{max}}\left(\cE_{\lambda,\sigma_{\mathrm{E}}}(\rho)\right) \leq g\big(\lambda N + (1-\lambda) N_{\mathrm{E}}\big)\ 
  \label{eq:maxout1}
\end{equation}
by the maximum entropy principle.
Combining Eqs.~\eqref{eq:upboundholevo}, \eqref{eq:minout1}, and \eqref{eq:maxout1}, the upper bound~\eqref{eq:upbound1} follows.\\

  \textbf{The lower bound.} 
  To show~\eqref{eq:lowbound1}, recall that taking the one-shot expression of the Holevo quantity and plugging in a specific ensemble of signal states $\{p_x, \rho_x\}_x$ gives a lower bound on the capacity. We pick a Gaussian ensemble of coherent states, $\{\frac{1}{2\pi N}e^{-\frac{|\xi|^2}{2N}}\mathrm{d}^2\xi, \proj{\xi}\}_\xi$. Note that the ensemble average~$\overline{\rho} = \frac{1}{2\pi N}\int e^{-\frac{|\xi|^2}{2N}} \proj{\xi} \mathrm{d}^2\xi$ is the Gaussian thermal state $\rho_{\mathrm{th},N}$ with mean photon number~$N$. Therefore  $S(\overline{\rho}) = g(N)$. A lower bound on the classical capacity is thus given by
  \begin{align}
    C_N(\cE_{\lambda,\sigma_{\mathrm{E}}}) &\geq \chi_N(\cE_{\lambda,\sigma_{\mathrm{E}}}) \\
    &\geq S\big(\cE_{\lambda,\sigma_{\mathrm{E}}}(\rho_{\mathrm{th},N})\big) - \frac{1}{2\pi N}\int e^{-\frac{|\xi|^2}{2N}} S\big(\cE_{\lambda,\sigma_{\mathrm{E}}}(\proj{\xi})\big) \mathrm{d}^2\xi\ . \label{eq:lowboundholevo1}
  \end{align}
We can lower bound the first term as follows: we have 
\begin{equation}
  S\big(\cE_{\lambda,\sigma_{\mathrm{E}}}(\rho_{\mathrm{th},N})\big) = S\big(\cE_{1-\lambda, \rho_{\mathrm{th},N}}(\sigma_{\mathrm{E}})\big) \geq g\big(\lambda N + (1-\lambda)N_\mathrm{E}^\mathrm{ep}\big)\ .
  \label{eq:minoutbeamsplit}
\end{equation}
The first equality follows because for general states $\rho, \sigma$, we have $\cE_{\lambda,\sigma}(\rho) = \cE_{1-\lambda,\rho}(\sigma)$, as can be seen by considering the characteristic function of the output state under a beamsplitter \cite{KoeSmiEPI}: it satisfies~$ \chi_{\cE_{\lambda,\sigma_{\mathrm{E}}}(\rho)}(\xi) = \chi_{\rho}(\sqrt{\lambda}\xi)\cdot \chi_{\sigma_{\mathrm{E}}}(\sqrt{1-\lambda}\xi) = \chi_{\cE_{1-\lambda},\rho(\sigma_{\mathrm{E}})}(\xi)$  for all $\xi\in\mathbb{R}^2$. The inequality in Eq.~\eqref{eq:minoutbeamsplit} follows from the lower bound~\eqref{eq:minout} 
on the output entropy of the phase-covariant Gaussian channel~$\cE_{1-\lambda,\rho_{\mathrm{th},N}}$ for fixed input entropy $S(\sigma_{\mathrm{E}})$.

To bound the second term in~\eqref{eq:lowboundholevo1}, observe that by the  maximum entropy principle, we have 
\begin{align}
  S(\cE_{\lambda,\sigma_{\mathrm{E}}}(\proj{\xi})) &\leq S([\cE_{\lambda,\sigma_{\mathrm{E}}}(\proj{\xi})])\\
  &= S\big(\cE_{\lambda,[\sigma_{\mathrm{E}}]}(\proj{\xi})\big)\\
  &= S\big(\cE_{\lambda,[\sigma_{\mathrm{E}}]}(\proj{0})\big)\\
  &\leq \sup_{\substack{\sigma_{\mathrm{E}}\;\mathrm{ Gaussian}\\ \tr(a_{\mathrm{E}}^\dagger a_{\mathrm{E}} \sigma_{\mathrm{E}})\leq N_{\mathrm{E}}}} S\big(\cE_{\lambda,\sigma_E}(\proj{0})\big) =: A(\lambda,N_E)\ .
\end{align}
The first identity holds because both expressions $[\cE_{\lambda,\sigma_{\mathrm{E}}}(\proj{\xi})]$ and $\cE_{\lambda,[\sigma_{\mathrm{E}}]}(\proj{\xi})$ define the same Gaussian state, as can be verified by computing the associated covariance matrices. The second identity is a consequence of the compatibility of the beamsplitter with displacements \cite[Lemma VI.1]{KoeSmiEPI} and invariance of the von Neumann entropy under unitaries.

It remains to find an upper bound on $A(\lambda, N_{\mathrm{E}})$. In order to find such an upper bound, we consider the mean photon number at the output and apply Eq. \eqref{eq:outmax}, obtaining
\begin{align}
  \tr\left(a^\dagger a \cE_{\lambda,\sigma_{\mathrm{E}}}(\proj{0})\right)
  &\leq (1-\lambda)N_{\mathrm{E}}\ .
\end{align}
Hence by the maximum entropy principle, we have that
\begin{equation}
  A(\lambda,N_{\mathrm{E}}) \leq g\big((1-\lambda)N_{\mathrm{E}}\big)\ .
\label{eq:lowbound2}
\end{equation}
Combining Eqs.~\eqref{eq:lowboundholevo1}, \eqref{eq:minoutbeamsplit}, and \eqref{eq:lowbound2}, the lower bound~\eqref{eq:lowbound1} follows.\\

\textbf{The difference between the upper and lower bound.}
The difference between the upper and lower bound is given by
\begin{equation}
\begin{aligned}
  \Delta(\cE_{\lambda,\sigma_{\mathrm{E}}})(N) = &\delta(N)+ g\big((1-\lambda) N_{\mathrm{E}}\big) - \log\left(\lambda + (1-\lambda) e^{S(\sigma_{\mathrm{E}})}\right)\ 
\end{aligned}
  \label{eq:diffbound1}
\end{equation}
where
\begin{equation}
  \delta(N)\ :=\ g\big(\lambda N + (1-\lambda) N_{\mathrm{E}}\big) - g\big(\lambda N + (1-\lambda) N_\mathrm{E}^\mathrm{ep}\big)\ 
\end{equation}
collects the terms depending on~$N$. Since $g'(N) = \log(N+1)-\log(N)$ is strictly decreasing, $N_\mathrm{E}^\mathrm{ep} \leq N_{\mathrm{E}}$ by the maximum entropy principle, and the monotonicity of $g$,  we have
\begin{equation}
  \delta'(N) = \lambda \bigg[g'\big(\lambda N+(1-\lambda)N_{\mathrm{E}}\big) - g'\big(\lambda N + (1-\lambda)N_\mathrm{E}^\mathrm{ep}\big)\bigg] \leq 0
\end{equation}
and $\delta$ is decreasing as well. Therefore, we have
\begin{equation}
  \delta(N) \leq \delta(0) = g\big((1-\lambda) N_{\mathrm{E}}\big) - g\big((1-\lambda)N_\mathrm{E}^\mathrm{ep}\big)\ .
\end{equation}
Inserting this into~\eqref{eq:diffbound1}, we finally obtain the bound~\eqref{eq:differenceboundone}, as claimed. 
\end{proof}

\begin{figure}[th]\centering
  \includegraphics[scale=0.5]{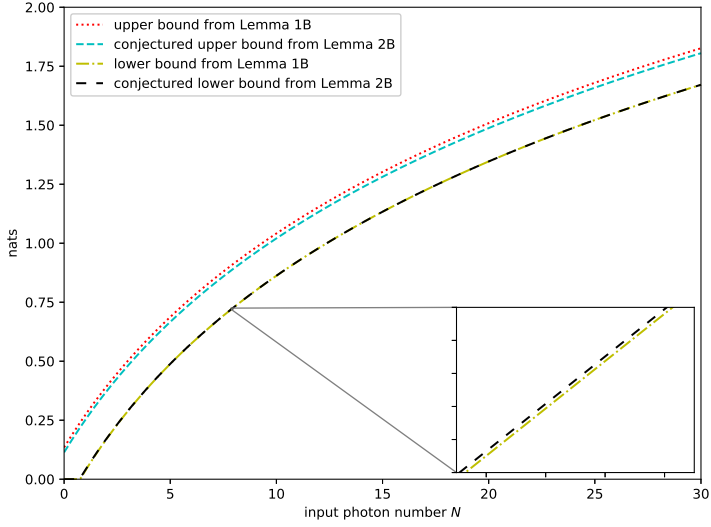}
  \caption{Bounds on the capacity $C_N(\cF_{t,f})$ with $t = 1$ for any distribution $f$ satisfying $\mathrm{E}(f) = 2$ and $e^{H(f)} \simeq 15.1 < 2\pi e$.} 
    \label{fig:2}
  \end{figure}

We now turn to the classical noise channel $\cF_{t,f}$ introduced in Eq. \eqref{eq:cqchannel}. The following Lemma immediately implies the second inequality in  Theorem~\ref{thm:main}. The bounds given in this lemma are illustrated in Fig.~\ref{fig:2}. 

\begin{customlemma}{1B}\label{lem:bound_additive1}
For the classical capacity of the single-mode classical noise channel $\cF_{t,f}$, we have the bounds \begin{align}
  C_N(\cF_{t,f}) &\geq \log\left(e^{g(N)} + t e^{H(f)}\right) - g(\pi t \mathrm{E}(f))\ ,\label{eq:lowbound1x}\\
  C_N(\cF_{t,f}) &\leq g\big(N+\pi t \mathrm{E}(f)\big) - \log\left(1 + t e^{H(f)}\right)\ .\label{eq:upbound1x}
\end{align}
The lower bound~\eqref{eq:lowbound1x} is achievable with a coherent state ensemble.  The difference between this upper and lower bound is bounded by
\begin{equation}
  \Delta(\cF_{t,f}) \leq 2g\big(\pi t \mathrm{E}(f)\big) - \log\left( 1 + t e^{H(f)}\right)\ ,
\end{equation}
independently of the input photon number $N$.
\end{customlemma}

\begin{proof}[Proof of Lemma~\ref{lem:bound_additive1}]
\textbf{The upper bound.}
To obtain the upper bound~\eqref{eq:upbound1x}, we again bound the Holevo quantity by
\begin{align}
  C_N(\cF_{t,f}) \leq S_N^{\mathrm{max}}(\cF_{t,f}) - \lim_{n \rightarrow \infty} \frac{1}{n} S^{\mathrm{min}}(\cF_{t,f}^{\otimes n})\ .\label{eq:holevoftf}
\end{align}
By the entropy power inequality~\eqref{eq:cqepi} we have 
\begin{align}
  \frac{1}{n} S^{\mathrm{min}}(\cF_{t,f}^{\otimes n}) \geq \log\left(1 + t e^{H(f)}\right)\ .
\end{align}
This is because $e^{S(\rho)/n}\geq 1$ for all $n$-mode states~$\rho$, 
and because
\begin{align}
  \cF_{t,f}^{\otimes n}(\rho_n) = \int \tilde{f}^{(n)}(\xi) W(\sqrt{t}\xi) \rho_n W(\sqrt{t}\xi)^\dagger \mathrm{d}^{2n}\xi
\end{align}
for the probability density function
\begin{align}
  \tilde{f}^{(n)}(\xi) = \Pi_{i=1}^n f(\xi_{2i-1},\xi_{2i}) = \Pi_{i=1}^n f(q_i,p_i)\qquad\textrm{ for }\xi=(\xi_1,\ldots,\xi_{2n})\in\mathbb{R}^n\ .
\end{align}
on $\mathbb{R}^{2n}$, which has Shannon entropy~$H(\tilde{f}^{(n)}) = n H(f).$

To bound the first term in~\eqref{eq:holevoftf}, we again use the maximum entropy principle to obtain 
\begin{align}
  S_N^{\mathrm{max}}(\cF_{t,f}) &= \sup_{\tr(a^\dagger a \rho) \leq N} S(\cF_{t,f} (\rho)) \\
  &\leq \sup_{\tr(a^\dagger a \rho) \leq N} S([\cF_{t,f}(\rho)]) \\
  &= \sup_{\tr(a^\dagger a \rho)\leq N} S\big(\cF_{t,[f]}([\rho])\big)\\
  &= g(N + \pi t \mathrm{E}(f))\ ,
\end{align}
giving the claimed upper bound. The last step follows because for Gaussian $f$ and $\rho$, it is easy to directly evaluate the behavior of the mean photon number under the channel and conclude that we can
replace $\sup_{\tr(a^\dagger a\rho) \leq N} S([\cF_{t,f} (\rho)])$ with $\sup_{\tr(a^\dagger a \rho) \leq N + \pi t\mathrm{E}(f)} S([\rho])$.\\

\textbf{The lower bound.} For the lower bound~\eqref{eq:lowbound1x} we use a Gaussian ensemble $\{g(\xi)\mathrm{d}^2\xi,\proj{\xi}\}_\xi$ (where $g(\xi)=\frac{1}{2\pi N} e^{-\frac{|\xi|^2}{2N}}$) of displaced coherent states with mean photon number (of the ensemble)~$N$. Then the ensemble average is~$\rho_{\mathrm{th},N} = \frac{1}{2\pi N} \int e^{-\frac{|\xi|^2}{2N}}\proj{\xi} \mathrm{d}^2\xi$ and we obtain
\begin{align}
  C_N(\cF_{t,f}) \geq \chi_N(\cF_{t,f})&\geq S\big(\cF_{t,f}(\rho_{\mathrm{th},N})\big) - \int  g(\xi) S\big(\cF_{t,f}(\proj{\xi})\big)\mathrm{d}^2\xi\\
  &\geq \log\left(e^{S(\rho_{\mathrm{th},N})} + t e^{H(f)} \right) - S\big(\cF_{t,f} (\proj{0})\big)\\
  &\geq \log\left(e^{g(N)} + t e^{H(f)}\right) - g\big(\pi t \mathrm{E}(f)\big)\ ,
\end{align}
where we have used the entropy power inequality~\eqref{eq:cqepi} and  invariance of the von Neumann entropy of a state under unitary conjugation, as well as compatibility of the classical noise channel with displacements \cite[Lemma 2]{HubKoeVersh16} in the first step and the fact that $\rho_{\mathrm{th},N}$ is a Gaussian thermal state with mean photon number $N$ in the second step.\\

\textbf{The difference between the upper and lower bounds.}
We again write the difference between the upper and lower bound as
\begin{align}
  \Delta(\cF_{t,f})(N)&=\delta(N)-\log\left(1+te^{H(f)}\right)+g\big(\pi t \mathrm{E}(f)\big)\ ,
\end{align}
where
\begin{align}
  \delta(N)=  g\big(N+\pi t \mathrm{E}(f)\big)-\log\left(e^{g(N)}+te^{H(f)}\right)\ .
\end{align}
We have
\begin{align}
  \log\left( e^{g(N)} + te^{H(f)}\right) &= \log\bigg[e^{g(N)}\left(1 + te^{H(f)-g(N)}\right) \bigg] \\
  &= g(N) + \log\left(1+t e^{H(f)-g(N)}\right) \geq g(N)\ ,
\end{align}
and thus
\begin{align}
  \delta(N)\leq  g\big(N+\pi t\mathrm{E}(f)\big)-g(N)\ .
\end{align}
Now an adaptation of the argument given in the proof of Lemma~\ref{lem:beamsplitterunconditional} proves the claimed statement: the rhs.~is monotonically decreasing in $N$, and thus maximal for $N=0$: this yields $\delta(N)\leq g\big(\pi t\mathrm{E}(f)\big)$, which implies the claim.
\end{proof}

\subsection{Derivation of strengthened capacity bounds from conjectured EPNIs}
We now sketch how to obtain Theorem~\ref{thm:bound_additive}, focusing only on the differences in the derivation between Theorem~\ref{thm:main} and Theorem~\ref{thm:bound_additive}. Again, we first treat the case of the attenuation channel~$\cE_{\lambda,\sigma_{\mathrm{E}}}$. The classical noise channel~$\cF_{t,f}$ is then discussed in Lemma~\ref{lem:bound_additive2} below. 

\begin{customlemma}{2A}\label{lem:bound_beamsplit2}
  Assuming the EPNI conjecture \eqref{eq:epni} holds, the classical capacity of $\cE_{\lambda,\sigma_{\mathrm{E}}}$ satisfies
\begin{align}
  C_N(\cE_{\lambda,\sigma_{\mathrm{E}}}) &\leq g\big(\lambda N + (1-\lambda) N_{\mathrm{E}} \big) - g\big( (1-\lambda) N_\mathrm{E}^\mathrm{ep}\big)\ .
\end{align}
The difference between this upper bound and the lower bound \eqref{eq:lowbound1} is bounded by
\begin{align}
  \Delta (\cE_{\lambda,\sigma_{\mathrm{E}}}) \leq 2\bigg[g\big( (1-\lambda) N_{\mathrm{E}}\big) - g\big( (1-\lambda) N_\mathrm{E}^\mathrm{ep}\big) \bigg]\ ,
\end{align}
independently of the input photon number $N$.
\end{customlemma}
We note that the EPNI does not improve upon the lower bound on the classical capacity of $\cE_{\lambda,\sigma_{\mathrm{E}}}$ given in Lemma~\ref{lem:beamsplitterunconditional}. This is because the lower bound only requires a bound on the minimal output entropy of $\cE_{\lambda, \sigma_{\mathrm{E}}}$ for thermal input states, a case where the statement of the EPNI is already covered by the established Eq.~\eqref{eq:minout}.

\begin{proof}Lemma~\ref{lem:bound_beamsplit2} follows when we use the EPNI \eqref{eq:epni} instead of the EPI \eqref{eq:epi} in the proof. In particular, we obtain from the EPNI that
\begin{align}
  \frac{1}{n} S\big(\cE^{\otimes n}_{\lambda, \sigma_{\mathrm{E}}}(\rho_n)\big) &\geq g\bigg(\lambda g^{-1}\big[S(\rho_n)/n\big] + (1-\lambda) g^{-1}\big[S(\sigma_{\mathrm{E}}^{\otimes n})/n\big] \bigg) \\
  &\geq g\big((1-\lambda) N_\mathrm{E}^\mathrm{ep}\big)\ ,
\end{align}
since $S(\rho) \geq 0$ for any state~$\rho$. This expression replaces the second term in the upper bound of Lemma~\ref{lem:beamsplitterunconditional}. The bound on the difference between the upper and the lower bound in Lemma~\ref{lem:bound_beamsplit2} follows analogously to before.
\end{proof}

Employing the EPNI \eqref{eq:cqepni} instead of the EPI \eqref{eq:cqepi} in a similar fashion shows the second part of Theorem~\ref{thm:bound_additive} for the classical noise channel~$\cF_{t,f}$: 
\begin{customlemma}{2B}\label{lem:bound_additive2}
  Assuming Eq. \eqref{eq:minoutconj} holds, the classical capacity of $\cF_{t,f}$ satisfies
  \begin{align}
    C_N(\cF_{t,f}) &\geq g\left(N + \frac{t}{e} e^{H(f)}\right) - g\big(\pi t \mathrm{E}(f)\big)\ .
  \end{align}
  Assuming in addition  that the EPNI conjecture \eqref{eq:cqepni} holds, we further have
  \begin{align}
    C_N(\cF_{t,f}) &\leq g\big(N + \pi t \mathrm{E}(f)\big) - g\left(\frac{t}{e}e^{H(f)}\right)\ .
  \end{align}
The difference between this upper and lower bound is bounded by
\begin{align}
  \Delta(\cF_{t,f}) \leq 2\bigg[g\big(\pi t \mathrm{E}(f)\big) - g\left(\frac{t}{e}e^{H(f)}\right) \bigg] \ ,
\end{align}
independently of the input photon number $N$.
\end{customlemma}
\begin{proof}
 Here we obtain
\begin{align}
  \frac{1}{n} S^{\mathrm{min}}(\cF_{t,f}^{\otimes n}) \geq g\left(\frac{t}{e} e^{H(f)}\right)\ ,
\end{align}
replacing the second term in the upper bound of Lemma~\ref{lem:bound_additive1}. The first term in the lower bound is replaced by
\begin{align}
  S\big(\cF_{t,f}(\rho)\big) \geq g\left(N + \frac{t}{e} e^{H(f)}\right)\ ,
\end{align}
replacing the first term in the lower bound of the lemma. The maximal difference between the upper and the lower bound follows again analogously to the reasoning from before.
\end{proof}

\section{Discussion}
\label{sec:discussion}
We have derived bounds on the classical capacity of a general class of bosonic channels which includes both Gaussian as well as non-Gaussian examples. We have shown that for these channels, the rates achievable by modulation of coherent states are not too far away from the maximal achievable rate (using arbitrary, possibly entangled code states). The maximal gain resulting from the use of general coding  strategies is bounded by a channel-dependent constant independently of the average energy of the signal states. In particular, this means that these channels can only exhibit a limited amount of additivity violation: the product-state and general classical capacities essentially coincide. 

There are a few paths one could follow for future work: 
First, a proof of Eq. \eqref{eq:minoutconj} would be desirable as a first step towards a proof of the classical-quantum EPNI \eqref{eq:cqepni}, implying better bounds on the classical capacity of the channels considered here. 
One may also wonder about similar statements for the multi-mode versions of the considered channels. The two main ingredients of our proofs, the maximum entropy principle and the entropy power inequality, both hold in the multi-mode case. Furthermore, minimal output entropy results analogous to \eqref{eq:minout} have been shown in the multi-mode setting~\cite{Holevo_2015}, essentially providing us with all tools required in the proof of our bounds. Therefore, similar bounds should hold.
Another interesting question concerns the implications of the EPI and EPNI to other capacities than the classical capacity, such as the entanglement-assisted classical capacity. The recently proven conditional version of the EPI has been shown to imply an upper bound on the entanglement-assisted capacity~\cite{depalmaconditional,koenigconditionalepi2015}, but a corresponding lower bound for general non-Gaussian channels is missing so far. 

\section*{Acknowledgments}
We thank the anonymous referees for their comments.
This work is supported by the Technische Universit\"at M\"unchen -- Institute for Advanced Study, funded by the German Excellence Initiative and the European Union Seventh Framework Programme under grant agreement no. 291763. The authors acknowledge additional support by DFG project no. K05430/1-1.


\end{document}